\documentclass[epsf]{aa}
\usepackage{natbib}
\usepackage{graphicx}

\begin{document}

\title{Is KT Mon a Classical Nova?}
\subtitle{}
\authorrunning{T. Kato and H. Yamaoka}
\titlerunning{Is KT Mon a Classical Nova?}

\author{Taichi Kato\inst{1}
        \and Hitoshi Yamaoka\inst{2}
}

\institute{
  Department of Astronomy, Kyoto University, Kyoto 606-8502, Japan
  \and Faculty of Science, Kyushu University, Fukuoka 810-8560, Japan
}

\offprints{Taichi Kato, \\ e-mail: tkato@kusastro.kyoto-u.ac.jp}

\date{Received / accepted }

\abstract{
   KT Mon has been considered as a classical nova in 1942 based on its
light curve and spectroscopy.  However, we noticed a unusual feature
for a nova soon after its maximum: the early presence of high excitation
emission lines such as He\textsc{II} and N\textsc{III}.
We propose previously overlooked interpretations that KT Mon can be either
a WZ Sge-type dwarf nova and an X-ray transient, although the possibility
of an unusual or a recurrent nova is not completely excluded.
If KT Mon is a WZ Sge-type dwarf nova, the system is expected to have
a brown dwarf secondary.
If KT Mon is an X-ray transient, the system is a good candidate for the
nearest black-hole binary.  Within our knowledge, the observed features
seem to more strongly support the latter possibility.
In all possibilities, we can expect a recurrent outburst in meaningfully
near future.
}

\maketitle

\section{Introduction}

   KT Mon (Nova Mon 1942) was discovered by A. N. Vyssotsky on a Harvard
plate taken on 1943 January 2.  The object was first recorded
at m$_{\rm pg}$ = 10.3 on 1942 December 30.  \citet{gap54ktmon} presented
a light curve and a description of a spectrum taken by Vyssotsky
on the discovery plate.  According to \citet{gap54ktmon}, the object
rather rapidly faded (2 mag in 27 d).  The object was fainter
than m$_{\rm pg}$ = 13.35, 13.9 d prior to the first positive record on
1942 December 30.  \citet{gap54ktmon}
described the low-resolution spectrum by Vyssotsky as ``broad bright
lines H$\beta$, H$\gamma$, H$\delta$, 4686 He\textsc{II} and
4640 N\textsc{III}".  From these findings, \citet{gap54ktmon} concluded
that KT Mon is a nova.  \citet{due87novaatlas} classified the object
as an established moderately fast nova.

   However, several authors have noted the peculiarity of KT Mon.
\citet{pay77novaproc} deduced the distances of past novae, and found
three stars (KT Mon, CG CMa and WX Cet) are more than 15 kpc distant
from the Galactic Center.  The two objects in this list, CG CMa
\citep{kat99cgcma,due99cgcma} and WX Cet
\citep{bai79wzsge,odo91wzsge,kat01wxcet} are now established to be
SU UMa-type dwarf novae (For a recent review of dwarf novae and SU UMa-type
dwarf novae, see \citet{osa96review} and \citet{war95suuma}, respectively.)
Regarding KT Mon, \citet{pay77novaproc} described that ``KT Mon: whose
spectrum observed by Vyssotsky, and whose light curve obtained by
\citet{gap54ktmon} place it unequivocally as a nova" and suggested that
``further study might find KT Mon to be recurrent" based on this unusual
distance estimate.  Similar long-distance estimates have been given in
\citet{war87CVabsmag,sha97novarate}.  No quiescent counterpart has been
reported down to mag 20 \citep{due87novaatlas}.

\section{Overlooked Unusual Features of KT Mon}

   In addition to the unusually long distance estimate \citep{pay77novaproc},
which originally questioned the classical nova-nature of KT Mon, we further
noted a previously overlooked peculiarity from the available material.

   The early presence of high excitation lines, particularly
He\textsc{II} $\lambda$4686 and N\textsc{III} $\lambda$4640, is unusual
for typical development of nova spectra
\citep{GalacticNovae,ClassicalNovae,wil92novaspec}, since strong appearance
of N\textsc{III} $\lambda$4640 Bowen brend requires high-energy photons
\citep{vanpar95XBreview}.  Such a condition is hard to achieve during the
early stage of a classical nova outburst when thermal emissions from a
lower ($\sim 10^4$ K) temperature expanded atmosphere of the white dwarf
dominates.  High-excitation emission lines can appear when the contracting
nova (pseudo)photosphere reaches $T_{\rm eff} \sim 2.5\times 10^4$ K
(see e.g. \cite{ClassicalNovae} Chap. 5), which then produces sufficient
high-energy UV photons.  This condition is expected to be achieved at
$\sim$2.5 mag below the optical maximum, which agrees with the general
observational records \citep{wil94novaspec}.  From an examination of
photographic records of past nova spectra, these lines were reported to
appear at around 4 mag below the optical maximum.
Only a few very fast novae (e.g. V2487 Oph, \cite{fil98v2487ophiauc})
and fast recurrent novae (e.g. U Sco, Sect. \ref{sec:RN}) are known
to exhibit these high-excitation lines soon after their maxima, which are
presumably caused by an unusually thin ejecta.  Since the decline rate of
KT Mon was not particularly fast, this object apparently did not achieve
this condition at the time of Vyssotsky's observation.

    A relatively large ($\sim$1.5) color index (blue-red) inferred by
\citet{gap54ktmon} also needs to be reexamined.
Taking into the recent observation of H$\alpha$ flux of post-outburst
novae \citep{cia90m31nova} into account, the contribution of the H$\alpha$
line to the $R$ band is estimated to $\sim$1.0 mag.  Although much of the
originally reported color index was tentativity attributed to interstellar
reddening in interpreting KT Mon as being a classical nova
\citep{pay77novaproc,sha97novarate}, a more recent estimate of the
maximum reddening in this direction
$E(B-V)$ = 0.545 or $A_V$ = 1.806 \citep{sch98reddening} suggests
that the observed color more strongly reflects the contribution of the
H$\alpha$ emission line.  If the ejecta was indeed unusually thin
as in recurrent novae, the contribution of the H$\alpha$ line, however,
might be insufficient (cf. \cite{sek88usco}) to explain the overall
color index.

\section{New Interpretations}

   In recent years, more categories of eruptive objects have been identified
to show high-excitation emission lines during their early stage of outbursts:
U Sco-type recurrent novae \citep{ros88usco,sek88usco,iij02uscospec},
WZ Sge-type dwarf novae \citep{bab02wzsgeletter,kuu02wzsge} and
X-ray transients \citep{tan96XNreview,vanpar95XBreview}.  Considering the
strong expected impact if KT Mon indeed belongs to any of these objects,
we further examine these possibilities.

   First of all, we have performed the prenova search with available
plate scans (the DSS1 red, the DSS2 red, blue, infrared)
and available catalogs in the VizieR service.
\citet{gap54ktmon} gave a position of nova as
6$^h$ 19$^m$ 58$^s$.8, +5$^{\circ}$ 29' 46'' (equinox 1900), which
precesses to 6$^h$ 25$^m$ 18$^s$.9, +5$^{\circ}$ 26' 28'' (J2000.0).
\citet{due87novaatlas} re-examined the Harvard plates and give the
position as 6$^h$ 25$^m$ 18$^s$.46, +5$^{\circ}$ 26' 31''.7.
\citet{due87novaatlas} further indicated that the identification by
\citet{kha71novaID} (who used the crude GCVS position for identification)
with a GSC-cataloged star having position end figures 19$^s$.69,
32".88 (GSC 2.2.1) is incorrect.  Taking this indication
into consideration, the position of \citet{due87novaatlas} must have
an accuracy better than 10".  We confirmed that there is no promising
optical, IR, and X-ray counterpart within 10" from the
\citet{due87novaatlas} position.  We have thus confirmed the safe upper
limit of the quiescent KT Mon to be $V$ = 20.

\subsection{Recurrent Nova?}\label{sec:RN}

   This possibility was originally considered by \citet{pay77novaproc},
but has been overlooked to date.  Since the outburst amplitude of KT Mon
is larger than 10, we can safely exclude the possibility of a recurrent nova
with a giant secondary \citep{anu99RN}.  The remaining possibilities
are a U Sco-like object or an IM Nor/CI Aql-like object
\citep{kis01ciaql,mat01ciaql,kat02imnor}.
The former class represents extremely rapidly evolving novae and the latter
slower ones.  The reported light curve of KT Mon \citep{gap54ktmon} (decline
by 2 mag in 27 d) more suggests the latter class.  However, the spectroscopic
feature of KT Mon (presence of the He\textsc{II} emission line) does not
resemble the early spectra of IM Nor and CI Aql (see
\citet{kis01ciaql,mat01ciaql}), which did not show high excitation lines.
Since the difference between these
classes can be primarily attributed to the mass of the white dwarf
\citep{hac00uscoburst,hac01ciaql}, there still remains the possibility
that KT Mon is essentially a U Sco-type object, with a slightly less
massive white dwarf.

   Based on modern calculations \citep{hac00uscoburst,hac00uscoqui},
the optical maximum of U Sco corresponds to $M_{\rm V}$ = $-$6.4$\pm$0.4.
By assuming $0 \leq A_V \leq 1.8$, and the same peak maximum $M_{\rm V}$
as in U Sco, the range of acceptable distance of KT Mon becomes
6$\leq d \leq$26 kpc.  This range can only
slightly reduce the Galactocentric distance questioned in
\citet{pay77novaproc}.  Even if the distance is acceptable, the upper limit
quiescent $M_{\rm V}$ = +3.3$\pm$0.4 is hard to accept for a recurrent
nova which requires a high accretion rate \citep{hac00uscoqui}.  If KT Mon
is indeed a recurrent nova, there is a need for a special unidentified
mechanism to reduce the quiescent luminosity.

\subsection{WZ Sge-Type Star?}

   It has been now widely demonstrated that some of WZ Sge-type dwarf novae
show strong He\textsc{II} and C\textsc{III}/N\textsc{III} emission lines
during their outbursts.  There properties are not inconsistent with the
existing observation of KT Mon.  Since KT Mon is 3.3 mag fainter than WZ Sge
at maximum, a reasonable upper limit of $d \leq$ 200 pc can be derived
from the recent distance estimate of WZ Sge (45 pc, J. Thorstensen (2001),
cited in \cite{ste01wzsgesecondary}).  If this interpretation is correct,
the quiescent upper limit of KT Mon corresponds to $M_{\rm V} = +13.5$.
This value is not inconsistent with a recent example of a very
faint WZ Sge-type object \citet{vantee99v592her}.  Although an argument
exists regarding the distance of V592 Her itself \citep{kat02v592her},
$M_{\rm V} = +13.5$ is not incompatible with a combination of a binary with
a cool white dwarf and a brown dwarf \citet{vantee99v592her}.  If KT Mon
belongs to WZ Sge-type dwarf novae, this object is a strong candidate for
a close binary containing a brown dwarf.

   This interpretation may seem to show a difficulty in interpreting the
reported large color index \citep{gap54ktmon} in its late-stage
light curve.  In recent years, however, WZ Sge-type dwarf novae can become
exceptionally red ($B-R \sim +1.0$ has been reported, see e.g.
\citet{pat98egcnc}) during the late stage of their outbursts.
This red color may explain some part of the color index reported by
\citet{gap54ktmon}.  WZ Sge-type dwarf novae are found to
frequently show ``rebrightenings".  In WZ Sge itself, such rebrightenings
with amplitudes of 1--2 mag \citep{ish02wzsgeletter}.  Since these
rebrightenings show a very rapid rise and fall, a scatter in the fragmentary
light curve in \citet{gap54ktmon} may reflect such a phenomenon.

   We also note that V4338 Sgr (Nova Sgr 1990), with similar spectroscopic
characteristics to those of KT Mon, has been classified as a possible
WZ Sge-type dwarf nova \citep{wag90v4338sgriauc}.

\subsection{X-ray Transient?}

   The presence of strong Balmer and He\textsc{II} and
C\textsc{III}/N\textsc{III} emission lines is also very characteristic
to X-ray transients (mostly black-hole candidates; see comprehensive
reviews \cite{vanpar95XBreview,tan95BHXN,whi95XBreview}).
It will be noteworthy that the X-ray transient V404 Cyg = GS 2023+338
\citep{wag91v404cyg,cas91v404cyg} had been considered to be a classical
nova based on its 1938 outburst observation \citep{due87novaatlas}.  KT Mon
therefore can be a X-ray transient.  In particular, the description of the
spectrum closely agrees to that of V404 Cyg \cite{wag91v404cyg}.
The decline rate (0.07 mag d$^{-1}$, see \citet{gap54ktmon}) is not
inconsistent with the statistics of optical properties of X-ray transients
\citep{che97BHXN}.  Since the optical maximum of KT Mon is 0.9 mag brighter
than that of the closest ``classical" X-ray transient V616 Mon
($d$ = 1.4 kpc, \cite{esi00gumusv616monHST}), this interpretation would
bring KT Mon to $d \sim$1 kpc.  This possibility makes KT Mon a candidate
for a closest black-hole binary.
Although a different distance can be acceptable considering
wide diversity of properties of X-ray transients
\citep{che97BHXN,mir01j1118}, the strong presence of He\textsc{II} and
C\textsc{III}/N\textsc{III} emission lines is more consistent with
a high-luminosity X-ray transient.  The distance $d \sim$1 kpc corresponds
to a quiescent upper limit of $M_{\rm V} = +10$.
This value is not inconsistent with a short-period system accreting at
a very low quiescent accretion rate
(cf. \citet{vanpar94LMXBabsmag,vanpar95XBreview}).  The reported color
index is mildly consistent with the reported reddening of V616 Mon
$E(B-V) = 0.35-0.9$, \cite{mar94v616mon}) combined with the redder color
during the late stage of outbursts of X-ray transients (e.g.
\cite{kin96gumusv518per}).

\section{Summary}

   KT Mon has been considered as a classical nova in 1942 based on its
light curve and spectroscopy.  However, we noticed a previously overlooked
unusual feature: the early presence of high excitation emission
lines such as He\textsc{II}.  We examine three new possibilities
(recurrent nova, WZ Sge-type dwarf nova, X-ray transient).  We have found
that the possibilities of a WZ Sge-type dwarf nova or an X-ray transient
are more consistent with the modern knowledge, although the possibility
of a recurrent nova (or a nova with an unusual spectroscopic evolution)
is not completely excluded.  If KT Mon is a WZ Sge-type
dwarf nova, the system should have a brown dwarf.  If KT Mon is an X-ray
transient, the system is a good candidate for the nearest black hole.
In all possibilities, we can expect a recurrent outburst in meaningfully
near future (10--100 yr depending on the classification).  Detailed
observations during the possible next outburst, as well as deep quiescent
observations, for this obscure object are strongly recommended.

\vskip 1.5mm

This work is partly supported by a grant-in aid [13640239 (TK),
14740131 (HY)] from the Japanese Ministry of Education, Culture, Sports,
Science and Technology.
This research has made use of the Digitized Sky Survey producted by STScI, 
the ESO Skycat tool, the VizieR catalogue access tool.


\begin{thebibliography}{50}
\expandafter\ifx\csname natexlab\endcsname\relax\def\natexlab#1{#1}\fi

\bibitem[{Anupama \& Mikolajewska(1999)}]{anu99RN}
Anupama, G.~C. \& Mikolajewska, J. 1999, \aap, 344, 177

\bibitem[{Baba {et~al.}(2002)Baba, Sadakane, Norimoto, Ayani, Ioroi, Matsumoto,
  Nogami, Makita, \& Kato}]{bab02wzsgeletter}
Baba, H., Sadakane, K., Norimoto, Y., {et~al.} 2002, \pasj, 54, L7

\bibitem[{Bailey(1979)}]{bai79wzsge}
Bailey, J. 1979, \mnras, 189, 41P

\bibitem[{Bode \& Evans(1989)}]{ClassicalNovae}
Bode, M.~F. \& Evans, A. 1989, Classical Novae (Alden Press: Oxford)

\bibitem[{Casares {et~al.}(1991)Casares, Charles, Jones, Rutten, \&
  Callanan}]{cas91v404cyg}
Casares, J., Charles, P.~A., Jones, D. H.~P., Rutten, R. G.~M., \& Callanan,
  P.~J. 1991, \mnras, 250, 712

\bibitem[{Chen {et~al.}(1997)Chen, Shrader, \& Livio}]{che97BHXN}
Chen, W., Shrader, C.~R., \& Livio, M. 1997, \apj, 491, 312

\bibitem[{Ciardullo {et~al.}(1990)Ciardullo, Shafter, Ford, Neill, Shara, \&
  Tomaney}]{cia90m31nova}
Ciardullo, R., Shafter, A.~W., Ford, H.~C., {et~al.} 1990, \apj, 356, 472

\bibitem[{Duerbeck(1987)}]{due87novaatlas}
Duerbeck, H.~W. 1987, \ssr, 45, 1

\bibitem[{Duerbeck {et~al.}(1999)Duerbeck, Schmeer, Knapen, \&
  Pollacco}]{due99cgcma}
Duerbeck, H.~W., Schmeer, P., Knapen, J.~H., \& Pollacco, D. 1999,
  Inf. Bull. Variable Stars, 4759

\bibitem[{Esin {et~al.}(2000)Esin, Kuulkers, McClintock, \&
  Narayan}]{esi00gumusv616monHST}
Esin, A.~A., Kuulkers, E., McClintock, J.~E., \& Narayan, R. 2000, \apj, 532,
  1069

\bibitem[{Filippenko {et~al.}(1998)Filippenko, Leonard, Modjaz, Eastman,
  Takamizawa, Hanzl, \& Kowalski}]{fil98v2487ophiauc}
Filippenko, A.~V., Leonard, D.~C., Modjaz, M., {et~al.} 1998, \iaucirc, 6943

\bibitem[{Gaposchkin(1954)}]{gap54ktmon}
Gaposchkin, S. 1954, \aj, 59, 199

\bibitem[{Hachisu \& Kato(2001)}]{hac01ciaql}
Hachisu, I. \& Kato, M. 2001, \apjl, 553, L161

\bibitem[{Hachisu {et~al.}(2000{\natexlab{a}})Hachisu, Kato, Kato, \&
  Matsumoto}]{hac00uscoburst}
Hachisu, I., Kato, M., Kato, T., \& Matsumoto, K. 2000{\natexlab{a}}, \apjl,
  528, L97

\bibitem[{Hachisu {et~al.}(2000{\natexlab{b}})Hachisu, Kato, Kato, Matsumoto,
  \& Nomoto}]{hac00uscoqui}
Hachisu, I., Kato, M., Kato, T., Matsumoto, K., \& Nomoto, K.
  2000{\natexlab{b}}, \apjl, 534, L189

\bibitem[{Iijima(2002)}]{iij02uscospec}
Iijima, T. 2002, \aap, 387, 1013

\bibitem[{Ishioka {et~al.}(2002)Ishioka, Uemura, Matsumoto, Ohashi, Kato, Masi,
  Novak, Pietz, Martin, Starkey, Kiyota, Oksanen, Moilanen, Cook, Kral, Hynek,
  Kolasa, Vanmunster, Richmond, Kern, Davis, Crabtree, Beaulieu, Davis,
  Aggleton, Gazeas, Niarchos, Yushchenko, Mallia, Fiaschi, Good, Boyd, Sano,
  Morikawa, Moriyama, Mennickent, Arenas, Ohshima, \&
  Watanabe}]{ish02wzsgeletter}
Ishioka, R., Uemura, M., Matsumoto, K., {et~al.} 2002, \aap, 381, L41

\bibitem[{Kato {et~al.}(2001)Kato, Matsumoto, Nogami, Morikawa, \&
  Kiyota}]{kat01wxcet}
Kato, T., Matsumoto, K., Nogami, D., Morikawa, K., \& Kiyota, S. 2001, \pasj,
  53, 893

\bibitem[{Kato {et~al.}(1999)Kato, Matsumoto, \& Stubbings}]{kat99cgcma}
Kato, T., Matsumoto, K., \& Stubbings, R. 1999, Inf. Bull. Variable Stars,
  4760

\bibitem[{Kato {et~al.}(2002{\natexlab{a}})Kato, Uemura, Matsumoto, Garradd,
  Masi, \& Yamaoka}]{kat02v592her}
Kato, T., Uemura, M., Matsumoto, K., {et~al.} 2002{\natexlab{a}}, \pasj,
  in press, (astro-ph/0209283)

\bibitem[{Kato {et~al.}(2002{\natexlab{b}})Kato, Yamaoka, Liller, \&
  Monard}]{kat02imnor}
Kato, T., Yamaoka, H., Liller, W., \& Monard, B. 2002{\natexlab{b}}, \aap, 391,
  L7

\bibitem[{Khatisov(1971)}]{kha71novaID}
Khatisov, A.~S. 1971, Byull. Abatsumani Astrofiz. Obs., 40, 13

\bibitem[{King {et~al.}(1996)King, Harrison, \& McNamara}]{kin96gumusv518per}
King, N.~L., Harrison, T.~E., \& McNamara, B.~J. 1996, \aj, 111, 1675

\bibitem[{Kiss {et~al.}(2001)Kiss, Thomson, Ogloza, Fur\'{e}sz, \&
  Szil\'{a}di}]{kis01ciaql}
Kiss, L.~L., Thomson, J.~R., Ogloza, W., Fur\'{e}sz, G., \& Szil\'{a}di, K.
  2001, \aap, 366, 858

\bibitem[{Kuulkers {et~al.}(2002)Kuulkers, Knigge, Steeghs, Wheatley, \&
  Long}]{kuu02wzsge}
Kuulkers, E., Knigge, C., Steeghs, D., Wheatley, P.~J., \& Long, K.~S. 2002, in
  The Physics of Cataclysmic Variables and Related Objects, ed. B.~T.
  G\"{a}nsicke, K.~Beuermann, \& K.~Reinsch (San Francisco: ASP), 443

\bibitem[{Marsh {et~al.}(1994)Marsh, Robinson, \& Wood}]{mar94v616mon}
Marsh, T.~R., Robinson, E.~L., \& Wood, J.~H. 1994, \mnras, 266, 137

\bibitem[{Matsumoto {et~al.}(2001)Matsumoto, Uemura, Kato, Kiyota, Ayani,
  Kawabata, Kr'{a}l, Havl\'{\i}k, Kolasa, Nov\'{a}k, \& Masi}]{mat01ciaql}
Matsumoto, K., Uemura, M., Kato, T., {et~al.} 2001, \aap, 378, 487

\bibitem[{Mirabel {et~al.}(2001)Mirabel, Dhawan, Mignani, Rodrigues, \&
  Guglielmetti}]{mir01j1118}
Mirabel, I.~F., Dhawan, V., Mignani, R.~P., Rodrigues, I., \& Guglielmetti, F.
  2001, \nat, 413, 139

\bibitem[{O'Donoghue {et~al.}(1991)O'Donoghue, Chen, Marang, Mittaz, Winkler,
  \& Warner}]{odo91wzsge}
O'Donoghue, D., Chen, A., Marang, F., {et~al.} 1991, \mnras, 250, 363

\bibitem[{Osaki(1996)}]{osa96review}
Osaki, Y. 1996, \pasp, 108, 39

\bibitem[{Patterson {et~al.}(1998)Patterson, Kemp, Skillman, Harvey, Shafter,
  Vanmunster, Jensen, Fried, Kiyota, Thorstensen, \& Taylor}]{pat98egcnc}
Patterson, J., Kemp, J., Skillman, D.~R., {et~al.} 1998, \pasp, 110, 1290

\bibitem[{Payne-Gaposchkin(1957)}]{GalacticNovae}
Payne-Gaposchkin, C. 1957, The Galactic Novae (North-Holland: Amsterdam)

\bibitem[{Payne-Gaposchkin(1977)}]{pay77novaproc}
Payne-Gaposchkin, C. 1977, in Novae and Related Stars, ed. M.~Friedjung
  (Dordrecht: D. Reidel Publishing Company), 3

\bibitem[{Rosino \& Iijima(1988)}]{ros88usco}
Rosino, L. \& Iijima, T. 1988, \aap, 201, 89

\bibitem[{Schlegel {et~al.}(1998)Schlegel, Finkbeiner, \&
  Davis}]{sch98reddening}
Schlegel, D.~J., Finkbeiner, D.~P., \& Davis, M. 1998, \apj, 500, 525

\bibitem[{Sekiguchi {et~al.}(1988)Sekiguchi, Feast, Whitelock, Overbeek,
  Wargau, \& Jones}]{sek88usco}
Sekiguchi, K., Feast, M.~W., Whitelock, P.~A., {et~al.} 1988, \mnras, 234, 281

\bibitem[{Shafter(1997)}]{sha97novarate}
Shafter, A.~W. 1997, \apj, 487, 226

\bibitem[{Steeghs {et~al.}(2001)Steeghs, Marsh, Knigge, Maxted, Kuulkers, \&
  Skidmore}]{ste01wzsgesecondary}
Steeghs, D., Marsh, T., Knigge, C., {et~al.} 2001, \apjl, 562, L145

\bibitem[{Tanaka \& Lewin(1995)}]{tan95BHXN}
Tanaka, Y. \& Lewin, W. H.~G. 1995, in X-Ray Binaries, ed. W.~H.~G. Lewin,
  J.~van Paradijs, \& van~den Heuvel. E. P.~J. (Cambridge: Cambridge
  University Press), 126

\bibitem[{Tanaka \& Shibazaki(1996)}]{tan96XNreview}
Tanaka, Y. \& Shibazaki, N. 1996, \araa, 34, 607

\bibitem[{van Paradijs \& McClintock(1994)}]{vanpar94LMXBabsmag}
van Paradijs, J. \& McClintock, J.~E. 1994, \aap, 290, 133

\bibitem[{van Paradijs \& McClintock(1995)}]{vanpar95XBreview}
van Paradijs, J. \& McClintock, J.~E. 1995, in X-Ray Binaries, ed. W.~H.~G.
  Lewin, J.~van Paradijs, \& van~den Heuvel. E. P.~J. (Cambridge: Cambridge
  University Press), 58

\bibitem[{van Teeseling {et~al.}(1999)van Teeseling, Hessman, \&
  Romani}]{vantee99v592her}
van Teeseling, A., Hessman, F.~V., \& Romani, R.~W. 1999, \aap, 342, L45

\bibitem[{Wagner {et~al.}(1991)Wagner, Bertram, Starrfield, Howell, Kreidl,
  Bus, Cassatella, \& Fried}]{wag91v404cyg}
Wagner, R.~M., Bertram, R., Starrfield, S.~G., {et~al.} 1991, \apj, 378, 293

\bibitem[{Wagner {et~al.}(1990)Wagner, Starrfield, \&
  Austin}]{wag90v4338sgriauc}
Wagner, R.~M., Starrfield, S.~G., \& Austin, S. 1990, \iaucirc, 5008

\bibitem[{Warner(1987)}]{war87CVabsmag}
Warner, B. 1987, \mnras, 227, 23

\bibitem[{Warner(1995)}]{war95suuma}
Warner, B. 1995, \apss, 226, 187

\bibitem[{White {et~al.}(1995)White, Nagase, \& Parmar}]{whi95XBreview}
White, N.~E., Nagase, F., \& Parmar, A.~N. 1995, in X-Ray Binaries, ed.
  W.~H.~G. Lewin, J.~van Paradijs, \& van~den Heuvel. E. P.~J.
  (Cambridge: Cambridge University Press), 1

\bibitem[{Williams(1992)}]{wil92novaspec}
Williams, R.~E. 1992, \aj, 104, 725

\bibitem[{Williams {et~al.}(1994)Williams, Phillips, \& Hamuy}]{wil94novaspec}
Williams, R.~E., Phillips, M.~M., \& Hamuy, M. 1994, \apjs, 90, 297

\end{thebibliography}
\end{document}